\documentstyle[graphicx,eqsecnum,epsfig,twocolumn,aps,floats]{revtex}

\begin{document}
\draft \preprint{}

\twocolumn[\hsize\textwidth\columnwidth\hsize\csname
@twocolumnfalse\endcsname

\title{Nature of the Electronic Excitations near the Brillouin Zone Boundary of
       Bi$_2$Sr$_2$CaCu$_2$O$_{8+\delta}$}
\author{D. L. Feng$^{1,2}$, C. Kim$^2$, H. Eisaki$^3$, D. H. Lu$^{2,3}$, K. M. Shen$^{2,3}$,
       F. Ronning$^{1,2}$, N. P. Armitage$^{1,2}$, A. Damascelli$^{2}$, N. Kaneko$^2$, M.
       Greven$^{2,3}$, J.-i. Shimoyama$^4$, K. Kishio$^4$, R. Yoshizaki$^5$, G. D. Gu$^6$ and Z.-X.Shen$^{1,2,3}$}
\address{$^1$Department of Physics, Stanford University, Stanford, CA 94305,
USA}
\address{$^2$ Stanford Synchrotron Radiation Laboratory, Stanford University, Stanford, CA 94305, USA}
\address{$^3$Department of Applied Physics, Stanford University, Stanford, CA 94305,
USA}
\address{$^4$Department of Applied Chemistry, University of Tokyo, Tokyo,
113-8656, Japan}
\address{$^5$Institute of Applied Physics, University of Tsukuba, Tsukuba, Ibaraki 305, Japan}
\address{$^6$School of Physics, University of New South Wales, P. O. Box 1,
Kensington, NSW, Australia 2033}

\date{\today}
\maketitle

\begin{abstract}

Based on angle resolved photoemission spectra measured on
different systems at different dopings, momenta and photon
energies, we show that the anomalously large spectral linewidth in
the $(\pi,0)$ region of optimal doped and underdoped
Bi$_2$Sr$_2$CaCu$_2$O$_{8+\delta}$ has significant contributions
from the bilayer splitting, and that the scattering rate in this
region is considerably smaller than previously estimated. This new
picture of the electronic excitation near $(\pi,0)$ puts
additional experimental constraints on various microscopic
theories and data analysis.

\end{abstract}
 \vskip2pc] 

\narrowtext

Angle resolved photoemission spectroscopy (ARPES) data from the
$(\pi,0)$ region of Bi$_2$Sr$_2$CaCu$_2$O$_{8+\delta}$ (Bi2212)
have been one of the most important sources of information about
the electronic structure of the high temperature superconductors
(HTSC)\cite{ARPESHTSC}. The normal state spectra are very broad,
with widths much larger than those from the nodal region (near
$(\pi/2,\pi/2)$), indicating a large anisotropy in the scattering
rate along the Fermi surface\cite{Shen97,Valla2}. This anisotropy
has been considered in various theories that describe the
anomalous transport and optical properties in the
cuprates\cite{Hotspot}. In addition, the information gathered from
these spectra have helped to put additional important parameters
into microscopic models\cite{Kim98}. On the other hand, the
superconducting state spectra contains the well known
peak-dip-hump structure\cite{Dessau91}. The position of the dip
was suggested to be related to the neutron $(\pi,\pi)$ resonance
mode\cite{Campuzano99}, resulting in modelling of the tunnelling
and ARPES data \cite{Eschrig00}. The peak intensity in the
peak-dip-hump structure has been interpreted as being related to
the condensate fraction\cite{Feng00}, as discussed by various
theories\cite{RVB,Stripe}. These studies constitute a significant
part of the HTSC literature.

Bi2212 has two coupled CuO$_2$ planes in the unit cell and
therefore bilayer splitting is naturally expected. However, it has
been largely ignored in the studies mentioned above, partly
because of earlier reports of its absence in the ARPES
spectra\cite{Ding96,Mesot01}. Recently, this long-sought bilayer
splitting was finally observed in overdoped
Bi2212\cite{Feng01a,BBS2,BBS3}. The two originally degenerate
bands (one for each CuO$_2$ layer) are split into bonding and
antibonding bands due to the intra-bilayer coupling. In the
$(\pi,0)$ region, the amplitude of the bilayer splitting is found
to be about 100 meV, comparable to the size of the superconducting
gap and the normal state band dispersion. As a result, the bilayer
splitting causes a peak-dip-hump structure even in the normal
state of heavily overdoped Bi2212\cite{Feng01a,Rast00,Zikris01},
demonstrating that the intra-bilayer coupling plays an important
role in the electronic structure of the overdoped regime and
should be seriously considered in relevant theories. These results
naturally raise the question of whether bilayer splitting exists
in the optimal and under-doped regimes where most experiments and
analyses were conducted, and if it does, how it affects our
understanding of the nature of the electronic excitations near
$(\pi,0)$.

In this letter, we report ARPES spectra from Bi2212 and
Bi$_2$Sr$_2$CuO$_{6+\delta}$ (Bi2201) for various dopings and
photon energies ($h\nu$).  The lineshapes of Bi2201 and Bi2212 are
similar in the nodal region, but very different near $(\pi,0)$. In
addition, Bi2212 spectra from the $(\pi,0)$ region are strongly
modified by $h\nu$, in contrast to the weak photon energy
dependence of the Bi2201 spectra. We show that these results can
be well explained by the underlying bilayer splitting effects in
under and optimally doped Bi2212 and that the broad linewidth near
$(\pi,0)$ is, in large part, due to the bilayer splitting. These
results are very different from the current, commonly-accepted
picture of the electronic excitations near $(\pi,0)$, and
therefore requires the reexamination of many existing theories,
and puts strong constraints on future theoretical models and data
analysis.

High quality Bi2212 and Bi2201 single crystals were grown by the
floating zone technique. Bi2212 samples are labeled by the
superconducting phase transition temperature $T_c$ of the sample
with the prefix UD for underdoped, OP for optimally doped, and OD
for overdoped. Bi2201 samples are labeled in the same way but in
lowercase. For example, UD83 represents a $T_c$=83\,K underdoped
Bi2212 sample, while od17 represents a $T_c$=17\,K overdoped
Bi2201 sample. Samples with Pb doping are labeled with the prefix
``Pb", except od33, which is doped with both Pb and La. The
superconducting transition widths, $\Delta T_c$, were less than
$3$K for all the samples used. Angle resolved photoemission
experiments were performed at a normal incidence monochromator
(NIM) beamline of the Stanford Synchrotron Radiation Laboratory,
where the intensity of the second order light is extremely weak.
Data were taken with a Scienta SES200 electron analyzer with the
angular resolution of $0.3\times 0.5$ degrees unless specified
otherwise. The overall energy resolution varied from 10 meV to 18
meV at different $h\nu$'s. This variation of the energy resolution
does not affect any of our conclusions since the energy scales of
the discussed features are much larger. The chamber pressure was
better than $5\times10^{-11} torr$, and sample aging effects were
negligible during the measurements. Unless otherwise specified,
normal state data were taken 10$\sim$20K above $T_c$.

\begin{figure}[t!]
\begin{center}
\epsfig{file=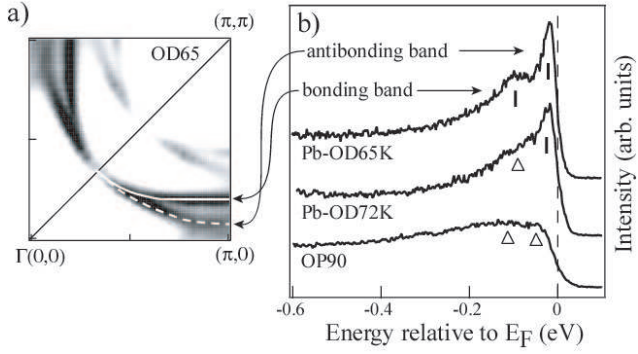,clip=,angle=0,width=3.3in}
\end{center} \vspace{0in}
 \caption{(a) Bilayer-split Fermi surfaces of
heavily overdoped OD65; the two weaker features are their
superstructure counter parts. Solid and dashed lines represent the
bonding and antibonding Fermi surfaces, respectively. (b) Normal
state photoemission spectra of Bi2212 taken at $(\pi,0)$ for three
different doping levels. Data were taken with $h\nu=22.7eV$
photon. Bars indicate identified feature positions, and triangles
indicate possible feature positions. }
 \label{autonum1}
\end{figure}

\begin{figure}[t]
\begin{center}
\epsfig{file=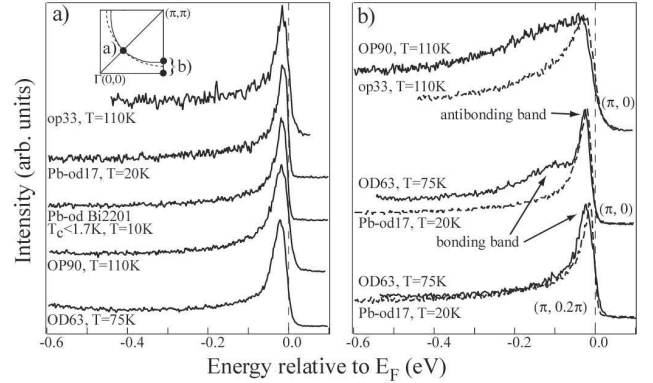,clip=,angle=0,width=3.3in}
\end{center}
 \vspace{0in}
 \caption{Normal state spectra taken at (a) the d-wave node region,
and (b) the $(\pi,0)$ region for both Bi2212 and Bi2201 at various
dopings. The solid and dashed lines in the inset of panel (a)
indicate the bonding and antibonding Fermi surfaces of OD63
respectively, and the black dots indicate the momentum locations
of the spectra. Data were taken with $h\nu=22.7 eV$. Angular
resolution was 0.3$^\circ$ for OP90 and 0.12$^\circ$ for others in
panel a).} \label{autonum2}
\end{figure}

The most obvious signature of the bilayer splitting is double
features in the Fermi surface and energy distribution curves
(EDC), as shown in Fig. 1 for heavily overdoped Bi2212
\cite{Feng01a}. Due to the anisotropic nature of the intrabilayer
coupling, the amplitude of the bilayer splitting is also
anisotropic with zero splitting in the nodal region and maximum
splitting at $(\pi,0)$ [Fig. 1(a)]. The normal state $(\pi,0)$ EDC
of heavily overdoped Pb-OD65 [Fig. 1(b)] clearly shows two
features that exhibit a normal state peak-dip-hump structure, and
are assigned to the two bilayer split
bands\cite{Feng01a,BBS2,BBS3}. This was not observed in previous
measurements on overdoped samples, mainly due to extrinsic factors
such as energy and angular resolution. With a slight decrease of
the doping (Pb-OD72), the two components of the $(\pi,0)$ spectrum
are barely distinguishable. Compared with Pb-OD65, the two
features become broader and their intensities smaller. For OP90,
the spectrum is intrinsically too broad to distinguish the two
split features, which makes the identification of the bilayer
splitting very difficult in this manner.

To clarify this further, we looked for other signs of bilayer
splitting by comparing the spectra of Bi2212 with those of single
layer Bi2201 at similar doping levels. We chose two pairs of
samples: OP90 and op33, and OD63 and od17. Based on the empirical
$T_c$ vs. doping formula\cite{Tallon}, they have doping levels of
0.16, 0.16, 0.22, and 0.24, respectively. For the spectra taken in
the nodal region shown in Fig. 2a, Bi2201 and Bi2212 have similar
lineshapes, and the linewidth varies only slightly for different
systems and experimental conditions. This holds true even for the
heavily overdoped Bi2201 sample with a $T_c\!<\!1.7K$ (doping
level $\sim$ 0.28). The situation is very different for the
spectra taken in the $(\pi,0)$ region (Fig. 2b). For OD63, the
spectrum consists of both the bonding and antibonding bands, while
the spectrum at $(\pi,0.2\pi)$ mostly consists of the bonding
band, because the antibonding band is above $E_F$ \cite{Feng01a}.
We find that the Bi2212 and Bi2201 spectra match at $(\pi,0.2\pi)$
almost perfectly, while those at at $(\pi,0)$ do not because of
the presence of the  bonding band at higher energies. As far as
the near-$E_F$ features are concerned, the spectra from both od17
and OD63 have very similar linewidths at similar binding energies
and momenta. This similarity between the OD63/od17 low energy
spectra can be attributed to their similar doping levels in each
CuO$_2$ plane. The OP90/op33 $(\pi,0)$ spectra show a large
mismatch similar to the OD63/od17 case, which can be naturally
attributed to the additional spectral weight from the bonding band
of OP90. On the other hand, without bilayer splitting (or
intra-bilayer coupling) in OP90, properties of the CuO$_2$ planes
of OP90 and op33 should be similar. It is then difficult to
explain why the linewidths of OP90 and op33 are so dramatically
different in the $(\pi,0)$ region, considering that Bi2201 and
Bi2212 are very similar in many other aspects such as the phase
diagram, Fermi surface shapes, dispersion energy
scales\cite{Jeff,Sato}, and particularly, residual resistivity,
which indicates the scattering caused by defects and impurities.
The larger linewidth of op33, compared to od17, may be attributed
to enhanced correlation effects with decreased doping, presumably
$(\pi,\pi)$ scattering due to increased antiferromagnetic
fluctuations\cite{Shen97,Hotspot}. We note that these normal state
spectra were taken at different temperatures. However, the thermal
broadening is negligible compared to the peak widths, within the
experimental temperature range.

\begin{figure}[t!]
\begin{center}
\epsfig{file=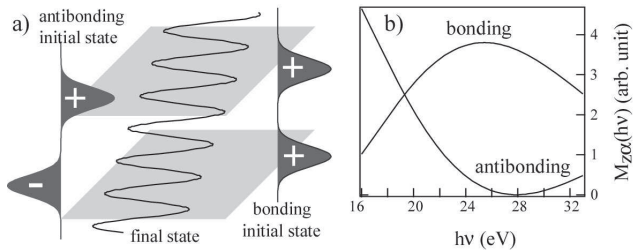,clip=,angle=0,width=3.3in}
\end{center}
 \vspace{0in}
 \caption{(a) Cartoon of the initial and final state
symmetries along the c-axis for the photoemission process with the
presence of the bilayer splitting. (b) The calculated c-axis
contributions to the photoemission matrix elements for both
bonding and antibonding state as functions of $h\nu$, as described
in the text. } \label{autonum3}
\end{figure}

The above comparison between spectra from Bi2201 and Bi2212
suggests the possible presence of bilayer splitting in optimally
doped samples. This is further supported by photon energy
dependence studies. As depicted in Fig. 3(a), the antibonding and
bonding states have opposite symmetry along the c-axis with
respect to the midpoint between the two $CuO_2$ planes. As a
consequence, their photoemission matrix elements respond
differently to various experimental parameters, including the
photon energy. Upon tuning $h\nu$, the spectral weight from the
bonding and antibonding states will vary differently, thus
changing the overall spectral lineshapes. This can be further
illustrated by an analysis of the photoemission matrix elements.
Although comprehensive calculations of the photoemission matrix
element are still not feasible because of the complexity in the
crystal structure and the photoemission process, as well as the
electron-electron correlations, with reasonable assumptions and
simplifications one can still study its behaviors on a qualitative
level, which turns out to be very helpful for the interpretation
of the data on various occasions\cite{Bansil99}.

\begin{figure*}[t!]
\begin{center}
\epsfig{file=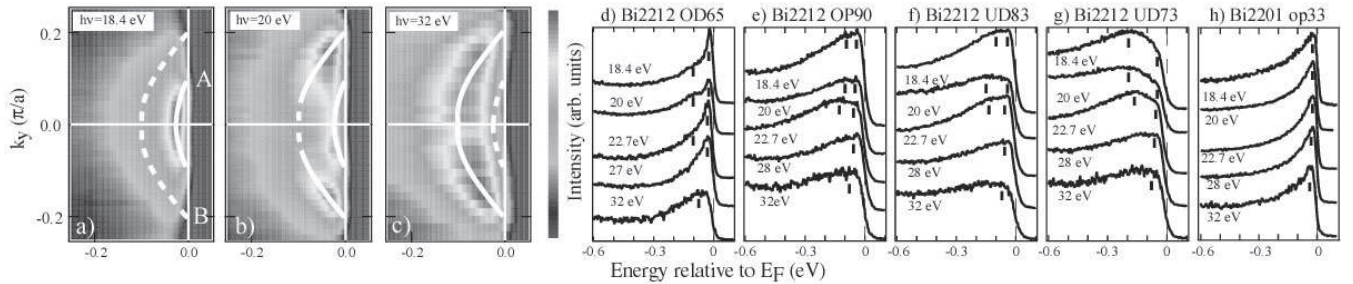,clip=,angle=0,width=7in}
\end{center}
 \vspace{0in}
 \caption{ (a-c) Normal state photoemission intensity as a function of
momentum and binding energy in grayscale maps near the $(\pi,0)$
region ($k_x=\pi$) of OD65 taken with $h\nu=$ 18.4 eV, 20 eV and
32 eV respectively. The thick white lines indicate the dispersion
of the bonding (B) and antibonding (A) bands, and lines are shown
in dashed form when the feature is weak. (d-h) Normal state EDC's
taken at $(\pi,0)$ for samples (d) OD65, (e) OP90, (f) UD83, (g)
UD73, and (h) op33. at various $h\nu$'s. Bars serve as guides for
the centroids of spectral features. Note that the features
indicated by bars near $E_F$ may be Fermi cutoffs instead of real
features. The separation between bars is not necessarily the
splitting energy. }\label{autonum4}
\end{figure*}

We consider two 2D systems coupled via a certain bilayer
interaction. The photoemission intensity for such a system with
non-interacting electrons is $I\propto \sum_{\alpha={a,b}}
M_{\alpha}^2 A_{\alpha}(\vec{k},\omega)$ where $\vec{k}$, $\omega$
and $A_{\alpha}$ are the momentum, energy for the final state, and
the spectral function respectively, while $a$ and $b$ represent
antibonding and bonding bands. In the one-electron matrix element
$ M_{\alpha} = \langle \psi_{f} | {\bf A \cdot p} | \psi_{i\alpha}
\rangle $, $\psi_f$ and $\psi_{i\alpha}$ are the final state and
initial state single electron wave functions, {\bf A} is the
vector potential of the photon field, and ${\bf
p}\equiv-i\hbar\nabla$. Assuming $
\psi_{i\alpha}(x,y,z)=\phi_i(x,y)\chi_{\alpha}(z) $ and a free
electron final state $\psi_f(x,y,z)=e^{ik_x x +ik_y y + i k_z z}$,
the matrix element can be separated into the in-plane contribution
$ M_{\parallel}$ and the out-of-plane contribution $ M_{z\alpha}
$. Under the dipole approximation,
\begin{eqnarray}
 M_{\alpha} & \propto & M_{\parallel}+ M_{z\alpha} \nonumber \\
 & \equiv &
     A_{\parallel}/A_z \langle e^{ik_x x +ik_y y} | r_{\parallel} | \phi_i(x,y) \rangle \
    +  \langle e^{ik_z z} | z | \chi_{\alpha}(z) \rangle \nonumber
\end{eqnarray}
where $k_x$ and $k_y$ are fixed to be $(\pi,0)$ for both the
initial and final states. The first term contributes equally to
both bonding and antibonding states, and the ratio of
polarization, $A_{\parallel}/A_{z}$, is approximately constant in
the experiment. Therefore, we can focus on $M_{z\alpha}$ as a
function of $h\nu$. To further simplify, we assume
$$\chi_{\alpha}(z)=e^{-\frac{(x-l_0/2)^2}{(\beta l_0)^2}} \pm
e^{-\frac{(x+l_0/2)^2}{ (\beta l_0)^2}},$$ where ``-" and ``+"
signs are for $\alpha=a$ and $b$, respectively. $l_0$ is the
intrabilayer distance, and $\beta$ is an adjustable parameter
reflecting how the electron wavefunction is localized within a
$CuO_2$ layer and is assumed to be $\beta=1/6$ in the calculation.
For the final state, the free electron approximation gives $k_z=
[2 m^*\hbar^{-2} ( h\nu - \Phi +v_0) -(k_x^2+k_y^2)
]^{-\frac{1}{2}}$, where we choose the photoelectron effective
mass $m^*$ to be the free electron mass, the work function
$\Phi=\,4.3\,eV$, and inner potential $v_0=\,7\,eV$ in the
calculation\cite{calcnote}. $M_{z\alpha}$ calculated with these
parameters and simplifications is shown for both antibonding and
bonding states in Fig. 3(b). The $M_{z\alpha}$'s for the bonding
and antibonding state have almost opposite behaviors with $h\nu$,
and changes quite dramatically in the studied $h\nu$ range. This
causes the overall lineshape of the Bi2212 $(\pi,0)$ spectrum to
alter significantly with $h\nu$ as the relative weight of
bonding/antibonding states oscillates. In the case of optimally
doped and underdoped systems, the centroid of the broad feature
will shift.

This is indeed observed in OD65, where the bilayer splitting has
been clearly identified\cite{Feng01a}. Fig.\,4(a-c) show ARPES
intensity taken in the $(\pi,0)$ region at different $h\nu$'s as a
function of momentum and binding energy. Because the NIM gives
extremely weak second-order light, it is possible to directly
compare spectra taken at different $h\nu$'s. One can see that the
relative intensities of the antibonding band (A) and bonding band
(B) change with $h\nu$. At some photon energies, only one feature
is prominent, while in others, both features are clearly visible.
EDC's of OD65 at $(\pi,0)$ are plotted in Fig. 4(d)\cite{Expnote}.
While some $k_z$ dispersion may exist, the data show strong
bilayer matrix element effects. One clearly sees that the relative
intensities of the bonding and antibonding features vary
drastically with $h\nu$. For optimally doped [Fig. 4(e)] and
underdoped Bi2212 [Fig. 4(f-g)], one does not see two clearly
separated features. However, one can see the strong variation of
the lineshape, and changes in the centroid of the feature.
Although there are some detailed variations from sample to sample,
the spectra of underdoped and optimally doped Bi2212 change with a
similar trend as the OD65. On the other hand, for the optimally
doped single layer system Bi2201 [Fig. 4(h)], the peak position
and the overall lineshape show virtually no photon energy
dependence. The high binding energy background of op33 is a smooth
function of binding energy and photon energy. These indicate that
the strong photon energy dependence of the Bi2212 spectra is due
to the bilayer splitting.

Because of the previous lack of evidence for bilayer splitting in
optimally doped and underdoped Bi2212 \cite{Ding96,Mesot01}, many
analyses and calculations assumed its absence. For example, the
momentum distribution curves in this region were usually fitted by
one Lorentzian\cite{Valla2,Bogdanov00}, when in fact it consisted
of two Lorentzians separated by the bilayer splitting in momentum
space. Bi2212 $(\pi,0)$ spectra were discussed
\cite{Shen97,Campuzano99}, and particularly, fitted with a
one-component formula\cite{Singlecomponent}. We show that even
with a two-component model, there are various uncertainties
involved in fitting the spectra, because the bonding and
antibonding features are weighted by unknown factors at certain
$h\nu$'s, and generally too broad to be reliably separated. For
overdoped Bi2212, the dip position will be shifted by the
overlapping of the two humps and the the superconducting peaks.

Intra-bilayer coupling was assumed in some theories to explain the
different temperature dependence behavior of c-axis and in-plane
transport and optical properties of bilayer systems
\cite{Ioffe98}. Our results reinforce the assumptions of these
theories. On the other hand, we find that the quasiparticle in the
$(\pi,0)$ region of optimally doped Bi2212 should be similar to
that of Bi2201, and thus much better defined than previously
believed from earlier Bi2212 data. The quasiparticle lifetime is
more than 100\% longer than obtained from previous EDC analyses,
{\sl i.e.}, the scattering rate in this $(\pi,0)$ ``hot spot" is
not as large as previously believed, although an anisotropy of the
scattering rate still exists in the optimally doped and underdoped
regime, as is observed in Bi2201\cite{Shen97,Hotspot}.

SSRL is operated by the DOE Office of Basic Energy Science
Divisions of Chemical Sciences and Material Sciences. The Material
Sciences Division also provided support for the work. The Stanford
experiments are also supported by the NSF grant DMR0071897 and ONR
grant N00014-98-1-0195-A00002. The crystal growth of op33,
Pb-od17, and Pb-OD65 at Stanford was supported by DOE under
Contract Nos. DE-FG03-99ER45773-A001 and DE-AC03-76SF00515. M.G.
is also supported by the A. P. Sloan Foundation.


\begin{references}
\bibitem{calcnote} We note that the qualitative results discussed
here are independent of the chosen parameters, and only the
qualitative results shall be compared with the data.
\bibitem{Expnote}
Data in Fig.4 (d-h) were taken in the momentum window of
$(0.01\!\sim\!0.02)\pi \times(0.02\!\sim\!0.03)\pi$ centered
around $(\pi,0)$. Due to the weak dispersion of the electronic
states near $(\pi,0)$ in Bi2212 and Bi2201, small variations in
the momentum window do not affect the spectral lineshape.
\end{references}
\end{document}